\documentclass[aps,prd,superscriptaddress]{revtex4}
\usepackage{amsmath}
\usepackage{epsfig}
\usepackage{float}
\newcommand{\bx}{\boldsymbol{x}}
\newcommand{\bz}{\boldsymbol{z}}
\newcommand{\bb}{\boldsymbol{b}}
\newcommand{\bk}{\boldsymbol{k}}
\newcommand{\bp}{\boldsymbol{p}}
\newcommand{\br}{\boldsymbol{r}}
\newcommand{\rmi}{\mathrm{i}}
\newcommand{\rmd}{\mathrm{d}}
\allowdisplaybreaks[4]

\begin{document}
\title{Multiparticle azimuthal angular correlations in $pA$ collisions}

\author{Cheng Zhang}\email{zhangcheng@mails.ccnu.edu.cn}
\affiliation{Key Laboratory of Quark and Lepton Physics (MOE) and Institute of Particle Physics, Central China Normal University, Wuhan 430079, China}

\author{Manyika Kabuswa Davy}\email{davymanyika@mails.ccnu.edu.cn}
\affiliation{Key Laboratory of Quark and Lepton Physics (MOE) and Institute of Particle Physics, Central China Normal University, Wuhan 430079, China}

\author{Yu Shi}\email{physhiyu@mails.ccnu.edu.cn}
\affiliation{Key Laboratory of Quark and Lepton Physics (MOE) and Institute of Particle Physics, Central China Normal University, Wuhan 430079, China}

\author{Enke Wang}\email{wangek@mail.ccnu.edu.cn}
\affiliation{Key Laboratory of Quark and Lepton Physics (MOE) and Institute of Particle Physics, Central China Normal University, Wuhan 430079, China}
\affiliation{School of Physics and Telecommunication Engineering, South China Normal University, Guangzhou 510006, China}

\begin{abstract}
\vspace{0.cm}
In the Color Glass Condensate formalism, we evaluate the 3-dipole correlator up to the $\frac{1}{N_c^4}$ order with $N_c$ being the number of colors, and compute the azimuthal cumulant $c_{123}$ for 3-particle productions. In addition, we discuss the patterns appearing in the $n$-dipole formula in terms of $\frac{1}{N_c}$ expansions. This allows us to conjecture the $N_c$ scaling of $c_n\{m\}$, which is crosschecked by our calculation of $c_2\{4\}$ in the dilute limit.
\end{abstract}
\maketitle

\section{Introduction}
Experimental data in deep inelastic scatterings (DIS) and proton-nucleus ($pA$) collisions indicates a sharp rise in the cross section at small Bjorken $x$, which is believed to be due to the fast growth of the gluon density in large nuclei. The gluon density increases steeply and tends to be saturated when going to a smaller $x$. The Color Glass Condensate (CGC) formalism \cite{Iancu:2002xk,Iancu:2003xm,Weigert:2005us,JalilianMarian:2005jf,arXiv:1002.0333,Kovchegov:2012mbw,Albacete:2014fwa} describes this gluon saturation regime by treating the fast-moving partons inside the dense nucleus as a classical color source of the soft gluons. In the McLerran-Venugopalan (MV) model \cite{McLerran:1993ni}, the small-$x$ gluons are assumed as a classical Yang-Mills field. The soft gluons, whose color field $A$ is associated with a random color source $\rho$, are radiated by the eikonal (high-energy and fixed at a transverse coordinate) partons \cite{Kovner:2001vi}.

With the fact that the gluon density inside the nucleus is large, one can obtain a weak coupling constant $g$. Nevertheless, due to the strong nuclear field $A\sim\frac{1}{g}$, the high-density regime is nonlinear and cannot be handled perturbatively. The Jalilian-Marian-Iancu-McLerran-Weigert-Leonidov-Kovner (JIMWLK) renormalization group equation \cite{Jalilian-Marian:1997jx+X,Ferreiro:2001qy} governs the nonlinear evolution of the gluon distribution function in the saturation regime. In the dilute regime, this equation reduces to the linear BFKL equation, which describes the evolution of the unintegrated gluon density.

Within the framework introduced above, multiparticle productions in high-energy collisions are represented by dipoles, quadrupoles, and higher-point functions. These multipoint functions are written in forms of the Gaussian average of Wilson lines in the fundamental representation for quarks and the adjoint representation for gluons. However, Ref. \cite{Dominguez:2012ad} shows that, in the large-$N_c$ limit, only dipoles and quadrupoles contribute to the particle production processes. The quadrupole corresponding to the Weisz\"acker-Williams gluon distribution has been evaluated in Ref. \cite{Dominguez:2010xd} by the JIMWLK Hamiltonian method.

The CGC formalism generates multiparticle correlations in dilute-dense scatterings ($pA$ collisions) from the initial state \cite{Armesto:2006bv,Dumitru:2008wn,Gavin:2008ev,Dumitru:2010mv,Dumitru:2010iy,Kovner:2010xk,Dusling:2012iga,Kovchegov:2012nd,Dumitru:2014dra,Dumitru:2014yza,Dumitru:2014vka,
Lappi:2015vha,Schenke:2015aqa,Lappi:2015vta,McLerran:2016snu,Kovner:2016jfp,Dusling:2017aot,Dusling:2017dqg,Fukushima:2017mko,Kovchegov:2018jun,Boer:2018vdi,Kovner:2018fxj,
Mace:2018vwq,Mace:2018yvl}. In order to study correlation observables, which include the $n$th moment of the $m$-particle cumulant $c_n\{m\}$ and the corresponding anisotropic flow $v_n\{m\}$, one has to derive multidipole correlators. Some recent work has been done in Refs. \cite{Fukushima:2017mko,Dusling:2017dqg,Dusling:2017aot,Kovner:2018vec}. In this paper, we evaluate the 3-dipole correlator up to the $\frac{1}{N_c^4}$ order by the color transition matrices method developed in Refs. \cite{HiroFujii,Blaizot:2004wv,Dominguez:2008aa,Dominguez:2011gc}. It helps to obtain an $n$-dipole formula in forms of a power series at large $N_c$. We calculate the 3-particle azimuthal cumulant $c_{123}$ \cite{Bhalerao:2011ry,Bhalerao:2011yg}, within the model calculation approach proposed in Refs. \cite{Lappi:2015vha,Lappi:2015vta,Dusling:2017dqg,Dusling:2017aot}. Finally, we discuss the $N_c$ scaling of $c_n\{m\}$, which is verified by the result of $c_2\{4\}$ in the dilute limit.

The rest of the paper is organized as follows. In Sec. \ref{Multi-dipole}, we briefly review the MV model, and evaluate the 2-dipole correlator in forms of large $N_c$ expansions. We calculate the 3-dipole correlator up to the $\frac{1}{N_c^4}$ order, then obtain the $n$-dipole expression. We subsequently proceed to compute the azimuthal cumulant $c_{123}$, and expand $c_2\{4\}$ at small saturation momentum scale $Q_s$ in Sec. \ref{Anisotropic}. We discuss the $N_c$ scaling of multiparticle correlations in Sec. \ref{scaling}. Finally, Sec. \ref{Conclusion} is for the summary.

\section{Multidipole Correlators in the McLerran-Venugopalan model}\label{Multi-dipole}
We aim to compute the $n$-dipole correlator, namely,
\begin{eqnarray}
\left<\prod\limits_{i=1}^nD(\bx_{2i-1\perp},\bx_{2i\perp})\right>\equiv\frac{1}{N_c^n}\left<\prod\limits_{i=1}^n\text{Tr}\left[U(\bx_{2i-1\perp})U(\bx_{2i\perp})^{\dagger}
\right]\right>,\label{ndipole}
\end{eqnarray}
where $U(\bx_\perp)$ is a Wilson line in the fundamental representation given by
\begin{equation}
 U(\bx_\perp)= \mathcal{P}\exp\Biggl[-\rmi g\int\limits_{-\infty}^{+\infty}\rmd x^+ A_a^-(x^+,\bx_\perp)t^a\Biggr],
\end{equation}
where $\mathcal{P}$ is $x^+$ ordering operator and $t^a$ is a color matrix in the fundamental $\textrm{SU}(N_\textrm{c})$ representation. In the MV model, $A_a^-$ is the classical color field that obeys the classical Yang-Mills equation
\begin{equation}
-\boldsymbol{\nabla}_\perp^2 A_a^-(x^+,\bx_\perp)=g\rho_a(x^+,\bx_\perp),\label{YM}
\end{equation}
where $\rho_a$ is the corresponding color charge density inside the nucleus. Deriving $A_a^-$ from Eq. (\ref{YM}) gives
\begin{equation}
A_a^-(x^+,\bx_\perp)=g\int\rmd^2 \bz_\perp G_0(\bx_\perp-\bz_\perp)\rho_a(x^+,\bz_\perp),\ \ G_0(\bx_\perp)=\int\frac{\rmd^2 \bk_\perp}{(2\pi)^2}\frac{e^{\rmi\bk_\perp\cdot\bx_\perp}}{\bk_\perp^2},
\end{equation}
where $G_0$ is the two-dimensional massless propagator.

The angle brackets in Eq. (\ref{ndipole}) represent the color field average of a given physical quantity $f[A]$, where in the MV model, the average is a Gaussian weighted functional integral in form of
\begin{equation}
\langle f[A]\rangle=\int\mathcal{D}\rho\exp\left\{-\int\rmd x^+\rmd^2\bx_\perp\frac{\rho_a^2(x^+,\bx_\perp)}{2\mu^2(x^+)}\right\}f[A],
\end{equation}
where $\mu^2(x^+)$ is the variance of the Gaussian distribution of color field, which represents the squared color charge per unit transverse area at coordinate $x^+$. The $\mu^2(x^+)$ integration over $x^+$ is proportional to the saturation momentum square $Q_{\rm s}^2$.

With the help of Wick's theorem, any correlator of $\rho$'s or $A$'s can be obtained by the most elementary correlators
\begin{eqnarray}
\langle\rho_a(x_1^+,\bx_{1\perp})\rho_b(x_2^+,\bx_{2\perp})\rangle&=&\delta_{ab}\delta(x_1^+-x_2^+)\delta(\bx_{1\perp}-\bx_{2\perp})\mu^2(x_1^+),
\\
g^2\langle A_a^-(x_1^+,\bx_{1\perp})A_b^-(x_2^+,\bx_{2\perp})\rangle&=&\delta_{ab}\delta(x_1^+-x_2^+)\mu^2(x_1^+)g^4\int\rmd^2\bz_\perp G_0(\bx_{1\perp}-\bz_\perp)G_0(\bx_{2\perp}-\bz_\perp)\nonumber
\\
&=&\delta_{ab}\delta(x_1^+-x_2^+)\mu^2(x_1^+)L_{12},
\end{eqnarray}
where
\begin{eqnarray}
L_{ij}\equiv g^4\int\rmd^2\bz_\perp G_0(\bx_{i\perp}-\bz_\perp)G_0(\bx_{j\perp}-\bz_\perp).
\end{eqnarray}

By using the color transition matrices method developed in Refs. \cite{HiroFujii,Blaizot:2004wv,Dominguez:2008aa,Dominguez:2011gc}, the 2-dipole correlator can be given by
\begin{eqnarray}
\left<D(\bx_1,\bx_2)D(\bx_3,\bx_4)\right>&=&\frac{T_\text{2-dipole}}{N_c^2}\left(\begin{matrix}N_{c}^2&N_{c} \end{matrix}\right)\sum\limits_{n=0}^\infty\frac{M_{\text{2-dipole}}^n}{n!}\left(\begin{matrix}1\\0\end{matrix}\right)\label{DD1}
\\
&=&\frac{T_\text{2-dipole}}{N_c^2}\left(\begin{matrix}N_{c}^2&N_{c}\end{matrix}\right)e^{M_{\text{2-dipole}}}\left(\begin{matrix}1\\0\end{matrix}\right),\label{DD2}
\end{eqnarray}
where $T_\text{2-dipole}=e^{-\frac{C_F}{2}\mu^2\sum\limits_{i=1}^{4}L_{ii}}$ is the so-called tadpole contribution corresponding to the configurations where each gluon link attaches to a single Wilson line. The $M_{\text{2-dipole}}$ is the color transition matrix of the 2-dipole configuration, which is written in terms of color transition factors $F$'s and $L$'s as
\begin{equation}
M_{\text{2-dipole}}
=\mu^2\left(\begin{matrix}LF_{12,34}&\frac{1}{2}F_{1423}\\\frac{1}{2}F_{1243}&LF_{14,32}\end{matrix}\right)\equiv\left(\begin{matrix}m_1&m_4\\m_2&m_3\end{matrix}\right),\label{M2}
\end{equation}
where $L_{ij,kl}\equiv L_{ij}+L_{kl}$, $LF_{ij,kl}\equiv C_FL_{ij,kl}+\frac{1}{2N_c}F_{ijkl}$, $\mu^2\equiv\int\rmd x^{+}\mu^{2}\left(x^{+}\right)$, and the color transition factor $F_{ijkl}\equiv L_{ik}+L_{jl}-L_{il}-L_{jk}$.

The matrix exponential $e^{M_{\text{2-dipole}}}$ in Eq. (\ref{DD2}) is the difficult part to calculate, which has been derived in Ref. \cite{Dominguez:2008aa} by using the eigenvalue method. In this paper we calculate the $n$th power of $M_{\text{2-dipole}}$ in Eq. (\ref{DD1}), which gives us a platform to derive the 3-dipole correlator and conjecture a general $n$-dipole correlator in forms of $\frac{1}{N_c}$ power expansions.

The vector $\left(\begin{matrix}1\\0\end{matrix}\right)$ in Eq. (\ref{DD1}), which represents the initial 2-dipole configuration, picks out the first column of the $n$th power of $M_{\text{2-dipole}}$
\begin{equation}
M_{\text{2-dipole}}^n\left(\begin{matrix}1\\0\end{matrix}\right)=\left(\begin{matrix}m_1^n+\sum\limits_{j=1}^{[\frac{n}{2}]}m_2^j m_4^j\sum\limits_{i=0}^{n-2j}\binom{j}{i+j}\binom{j-1}{n-i-j-1}m_1^im_3^{n-2j-i}
\\\sum\limits_{j=0}^{[\frac{n-1}{2}]}m_2^{j+1}m_4^j\sum\limits_{i=0}^{n-2j-1}\binom{j}{i+j}\binom{j}{n-i-j-1}m_1^im_3^{n-2j-1-i}\end{matrix}\right),
\end{equation}
where $[\frac{n}{2}]\equiv\text{Floor}(\frac{n}{2})$, namely the rounding down function. Summing over $n$ with the factor $\frac{1}{n!}$ gives the matrix exponential
\begin{equation}
e^{M_{\text{2-dipole}}}\left(\begin{matrix}1\\0\end{matrix}\right)=\left(\begin{matrix}e^{m_1}\sum\limits_{j=0}^\infty m_2^j m_4^j\prod\limits_{i=1}^{2j} \int\limits_0^{\xi_{i-1}}\rmd\xi_i e^{(-1)^{i+1}\xi_i(m_3-m_1)}
\\e^{m_1}\sum\limits_{j=0}^\infty m_2^{j+1}m_4^j\prod\limits_{i=1}^{2j+1}\int\limits_0^{\xi_{i-1}}\rmd\xi_i e^{(-1)^{i+1}\xi_i(m_3-m_1)}\end{matrix}\right).\label{e^M2}
\end{equation}

Substituting Eq. (\ref{e^M2}) into Eq. (\ref{DD2}) gives the 2-dipole correlator:
\begin{eqnarray}
&&\left<D(\bx_1,\bx_2)D(\bx_3,\bx_4)\right>\nonumber
\\
&=&T_\text{2-dipole}e^{m_1}\sum\limits_{j=0}^\infty\bigg[\underset{\frac{1}{N_c^{2j}}\text{order}}{\underbrace{m_2^jm_4^j}}\prod\limits_{i=1}^{2j}\int\limits_0^{\xi_{i-1}}\rmd
\xi_ie^{(-1)^{i+1}\xi_i(m_3-m_1)}+\underset{\frac{1}{N_c^{2j+2}}\text{order}}{\underbrace{\frac{m_2^{j+1}m_4^j}{N_c}}}\prod\limits_{i=1}^{2j+1}\int\limits_0^{\xi_{i-1}}\rmd
\xi_ie^{(-1)^{i+1}\xi_i(m_3-m_1)}\bigg],\ \xi_0=1,\label{DDm}
\end{eqnarray}
where $m_{1,3}$ are of $N_c^0$ order, and $m_{2,4}$ are of $\frac{1}{N_c}$ order.

One can also write down the 3-dipole correlator by definition as
\begin{eqnarray}
\left<D(\bx_1,\bx_2)D(\bx_3,\bx_4)D(\bx_5,\bx_6)\right>=\frac{T_{\text{3-dipole}}}{N_c^3}\left(\begin{matrix}N_c^3&N_{c(1\times3)}^2&N_{c(1\times2)} \end{matrix}\right)\sum_{n=0}^\infty\frac{M_{\text{3-dipole}}^n}{n!}\left(\begin{matrix}1\\0_{(5\times1)}\end{matrix}\right),\label{DDD1}
\end{eqnarray}
where $M_{\text{3-dipole}}$ is the corresponding color transition matrix given by
\begin{equation}
M_{\text{3-dipole}}=\mu^2\left(\begin{matrix}LF_{12,34,56}&\frac{1}{2}F_{1423}&\frac{1}{2}F_{1625}&\frac{1}{2}F_{3645}&0&0
\\\frac{1}{2}F_{1243}&LF_{14,32,56}&0&0&\frac{1}{2}F_{1645}&\frac{1}{2}F_{3625}
\\\frac{1}{2}F_{1265}&0&LF_{16,34,52}&0&\frac{1}{2}F_{3245}&\frac{1}{2}F_{1463}
\\\frac{1}{2}F_{3465}&0&0&LF_{12,36,54}&\frac{1}{2}F_{1623}&\frac{1}{2}F_{1425}
\\0&\frac{1}{2}F_{1465}&\frac{1}{2}F_{3425}&\frac{1}{2}F_{1263}&LF_{16,32,54}&0
\\0&\frac{1}{2}F_{3265}&\frac{1}{2}F_{1643}&\frac{1}{2}F_{1245}&0&LF_{14,36,52}\end{matrix}\right)
\equiv\left(\begin{matrix}M_{11(1\times1)}&M_{12(1\times3)}&0\\M_{21(3\times1)}&M_{22(3\times3)}&M_{23(3\times2)}\\0&M_{32(2\times3)}&M_{33(2\times2)} \end{matrix}\right),\label{M3}
\end{equation}
where $LF_{ab,cd,ef}\equiv C_F(L_{ab}+L_{cd}+L_{ef})+\frac{1}{2N_c}(F_{abcd}+F_{abef}+F_{cdef})$. Substituting the block matrix in Eq. (\ref{M3}) into Eq. (\ref{DDD1}), one obtains the first column of the $n$th power of the $M_{\text{3-dipole}}$. In this computation, we only derive the result up to the $\frac{1}{N_c^4}$ order
\begin{eqnarray}
&&M_{\text{3-dipole}}^n\left(\begin{matrix}1\\0_{(5\times1)}\end{matrix}\right)\Bigg|_{\text{to the}\frac{1}{N_c^4}\text{order}}\nonumber
\\
=&&\left(\begin{matrix}M_{11}^n+\sum\limits_{i=0}^{n-2}M_{12}M_{22}^{n-2-i}M_{21}M_{11}^i\binom{1}{i+1}+\sum\limits_{j=0}^{n-4}\sum\limits_{i=0}^{n-4-j}M_{12}
M_{22}^{{n-4-j-i}}M_{21}M_{12}M_{22}^jM_{21}M_{11}^i\binom{2}{i+2}
\\
+\sum\limits_{k=0}^{n-4}\sum\limits_{j=0}^{n-4-k}\sum\limits_{i=0}^{{n-4-k-j}}M_{12}M_{22}^{{n-4-k-j-i}}M_{23}M_{33}^kM_{32}M_{22}^jM_{21}M_{11}^i\binom{1}{i+1}
\\
\sum\limits_{i=0}^{n-1}M_{22}^{n-1-i}M_{21}M_{11}^i+\sum\limits_{j=0}^{n-3}\sum\limits_{i=0}^{n-3-j}M_{22}^{n-3-j-i}M_{21}M_{12}M_{22}^jM_{21}M_{11}^i\binom{1}{i+1}
\\
+\sum\limits_{k=0}^{n-3}\sum\limits_{j=0}^{n-3-k}\sum\limits_{i=0}^{n-3-k-j}M_{22}^{n-4-k-j-i}M_{23}M_{33}^kM_{32}M_{22}^jM_{21}M_{11}^i
\\
\sum\limits_{j=0}^{n-2}\sum\limits_{i=0}^{n-2-j}M_{33}^{{n-2-j-i}}M_{32}M_{22}^jM_{21}M_{11}^i
\end{matrix}\right).\label{M3^n}
\end{eqnarray}

\begin{figure}[H]
\vskip0.0\linewidth
\centerline{
\includegraphics[width = .8\linewidth]{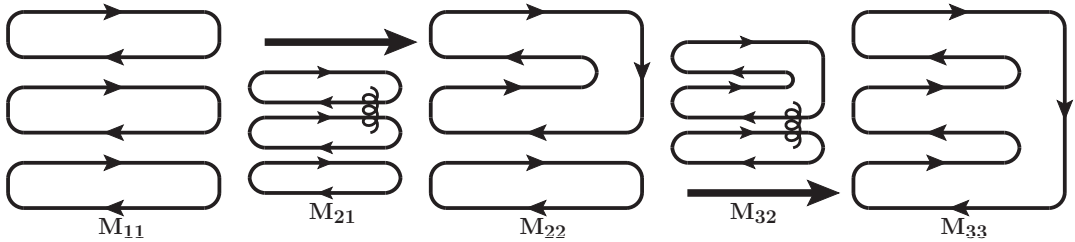}
}
\caption{The original configuration with three fermion loops goes through of color transitions two times, then it becomes another two configurations with two and one fermion loops, respectively.}
\label{1to2to3}
\end{figure}

Summing over $n$ with a factor $\frac{1}{n!}$ in Eq. (\ref{DDD1}), one obtains the $3$-dipole correlator up to the $\frac{1}{N_c^4}$ order
\begin{eqnarray}
&&\left<D(\bx_1,\bx_2)D(\bx_3,\bx_4)D(\bx_5,\bx_6)\right>\Bigg|_{\text{to the}\frac{1}{N_c^4}\text{order}}\nonumber
\\
&=&T_{\text{3-dipole}}\int\limits_0^1\prod_{i=0}^j\rmd\xi_i\delta(\sum_{i=0}^j\xi_i-1)\Bigg[\delta_{j0}+\delta_{j2}e^{\xi_2M_{11}}M_{12}e^{\xi_1M_{22}}M_{21} +\delta_{j4}e^{\xi_4M_{11}}M_{12}e^{\xi_3M_{22}}M_{21}e^{\xi_2M_{11}}M_{12}e^{\xi_1M_{22}}M_{21}\nonumber
\\
&&+\delta_{j4}e^{\xi_4M_{11}}M_{12}e^{\xi_3M_{22}}M_{23}e^{\xi_2M_{33}}M_{32}e^{\xi_1M_{22}}M_{21}+\frac{\delta_{j1}}{N_c}e^{\xi_1M_{22}}M_{21}+\frac{\delta_{j3}}{N_c} e^{\xi_3M_{22}}M_{21}e^{\xi_2M_{11}}M_{12}e^{\xi_1M_{22}}M_{21}\nonumber
\\
&&+\frac{\delta_{j3}}{N_c}e^{\xi_3M_{22}}M_{23}e^{\xi_2M_{33}}M_{32}e^{\xi_1M_{22}}M_{21}+\frac{\delta_{j2}}{N_c^2}e^{\xi_2M_{33}}M_{32}e^{\xi_1M_{22}}M_{21}\Bigg]
e^{\xi_0M_{11}},\label{DDDM}
\end{eqnarray}
where each term inside the square brackets has a clear physical pattern, which helps to conjecture the $n$-dipole correlator in the following sections. For illustration, we focus on the last term
\begin{eqnarray}
\frac{T_{\text{3-dipole}}}{N_c^2}\int\limits_0^1\rmd\xi_0\rmd\xi_1\rmd\xi_2\delta(\xi_0+\xi_1+\xi_2-1)e^{\xi_2M_{33}}M_{32}e^{\xi_1M_{22}}M_{21}e^{\xi_0M_{11}},
\end{eqnarray}
which indicates that the original configuration (whose topology is represented by $M_{11}$) goes through two times of color transitions (represented by $M_{21}$ and $M_{32}$), then becomes another two configurations (whose topologies are represented by $M_{22}$ and $M_{33}$). This term can be graphicly represented by the color transition approach below and in Fig. \ref{1to2to3}:
\begin{eqnarray}
M_{11}\overset{M_{21}}{\longrightarrow}M_{22}\overset{M_{32}}{\longrightarrow}M_{33},
\end{eqnarray}

Actually, the eight terms inside Eq. (\ref{DDDM}) correspond to eight kinds of color transition approaches, and all of them can be depicted as shown above.

On the other hand, one can check that each term of the multidipole correlators corresponds to an individual approach of color transitions between configurations, and vice versa. It enables us to conjecture an expression for the $n$-dipole correlator similar to the $2n$-point correlator \cite{Shi:2017gcq}, which reads
\begin{eqnarray}
\left<\prod_{i=1}^nD(\bx_{2i-1},\bx_{2i})\right>=T_{\text{$n$-dipole}}e^{M_1}\left[1+\sum_{k=1}^\infty\sum_{I_k}N_c^{L(U_{k+1}+U_0)-n}\prod_{i=1}^k \frac{\mu^2}{2}F_{a_ib_ic_id_i}\int\limits_0^{\xi_{i-1}}\rmd\xi_ie^{\xi_i\frac{\mu^2N_c}{2}F_{a_id_ic_ib_i}}\right],\label{nd}
\end{eqnarray}
where
\begin{eqnarray}
T_{\text{$n$-dipole}}=e^{-\frac{\mu^2C_F}{2}\sum\limits_{i=1}^{2n}L_{ii}},\ \
M_1=\mu^2(C_F\sum\limits_{i=1}^{n}L_{2i-1,2i}+\frac{1}{2N_c}\sum\limits_{I_1}F_{a_1b_1c_1d_1}),\ \ \xi_0=1.
\end{eqnarray}
$I_k$ represents summing over all possible permutations $a_1,b_1,c_1,d_1,a_2,b_2,c_2,d_2,...,a_{k},b_{k},c_{k},d_{k},$ which satisfy the following conditions
\begin{eqnarray}
&&a_i<d_i;\ \  (a_i,b_i),(d_i,c_i)\in U_i.
\end{eqnarray}
The recurrence relation between $U_i$ and $U_{i+1}$ is
\begin{eqnarray}
U_{i+1}=U_i-\{(a_i,b_i),(d_i,c_i)\}+\{(a_i,c_i),(d_i,b_i)\};\ \ i\geq1,
\end{eqnarray}
with its first two terms
\begin{eqnarray}
U_0=\{(2,1),(4,3),...,(2n,2n-1)\},\ \ U_1=\{(1,2),(3,4),...,(2n-1,2n)\}.
\end{eqnarray}
$U=\{(a,b),(b,c),...\}$, representing that the element $a$, is connected with $b$, and $b$ is connected with $c$,..., is a set of two-dimensional row matrices. $L(U)$ in Eq. (\ref{nd}) is a function of $U$, which equals the number of elements' loops in $U$. For example, if $U=\{(1,6),(2,3),(3,4),(4,5),(5,2),(6,1)\}$, which consists of two loops $1-6-1$ and $2-3-4-5-2$, then $L(U)=2$.

\section{Multi-particle azimuthal cumulants}\label{Anisotropic}
As the 3-dipole correlator involves such large numbers of color transitions, it is usually difficult to evaluate. As a first attempt, in this section, we calculate the 3-particle cumulant $c_{123}$ by using the 3-dipole correlator in Eq. (\ref{DDDM}) to show how it works. Generally, an $m$-particle correlation takes the form
\begin{eqnarray}
\langle e^{i(n_1\phi_1+n_2\phi_2+...+n_m\phi_m)}\rangle\equiv\frac{\kappa\{n_1,n_2,...,n_m\}}{\kappa\{0,0,...,0\}},
\end{eqnarray}
where $n_1,n_2,...,n_m$ are integers which satisfy the azimuthal symmetry: $n_1+n_2+...+n_m=0$. $\phi_i$ is the azimuthal angle of the $i$th outgoing particle's momentum $\bp_i$. $\kappa\{n_1,n_2,...,n_m\}$ is the mixed harmonic of the $m$-particle distribution, which is defined by
\begin{eqnarray}
\kappa\{n_1,n_2,...,n_m\}\equiv\int\prod_{i=1}^m\rmd^2\bp_ie^{in_i\phi_i}\frac{\rmd^mN}{\prod_{i=1}^m\rmd^2\bp_i},
\end{eqnarray}
where $\frac{\rmd^m N}{\prod_{i=1}^m\rmd^2\bp_i}$ is the $m$-particle inclusive spectra \cite{Lappi:2015vha,Lappi:2015vta,Dusling:2017dqg,Dusling:2017aot} given by
\begin{eqnarray}
\frac{\rmd^mN}{\prod_{i=1}^m\rmd^2\bp_i}&=&\int{\prod_{i=1}^m}\frac{\rmd^2\bb_i\rmd^2\br_i}{4\pi^3B_p}e^{-\frac{b_i^2}{B_p}-\frac{r_i^2}{4B_p}+i\bp_i\cdot \br_i}\left<\prod_{i=1}^mD\left(\bb_i+\frac{\br_i}{2},\bb_i-\frac{\br_i}{2}\right)\right>.
\end{eqnarray}

One of the nontrivial 3-particle cumulants is $c_{123}$, which is of the same definition with $v_{123}$ in Refs. \cite{Bhalerao:2011ry,Bhalerao:2011yg} and defined by
\begin{eqnarray}
c_{123}\equiv\frac{\kappa\{1,2,-3\}}{\kappa\{0,0,0\}}.\label{c3}
\end{eqnarray}
Within the framework above, the zeroth harmonic of the $m$-particle inclusive spectra in the MV model is derived by
\begin{eqnarray}
\kappa\{0,0,...,0\}&=&\int\prod_{i=1}^m\rmd^2\bp_i\frac{\rmd^m N}{\prod_{i=1}^m\rmd^2\bp_i}\nonumber
\\
&=&\int\prod_{i=1}^m\frac{\rmd^2\bb_i\rmd^2\br_i\rmd^2\bp_i}{4\pi^3B_p}e^{-\frac{b_i^2}{B_p}-\frac{r_i^2}{4B_p}+i\bp_i\cdot\br_i} \left<\prod_{i=1}^mD\left(\bb_i+\frac{\br_i}{2},\bb_i-\frac{\br_i}{2}\right)\right>\nonumber
\\
&=&\int\prod_{i=1}^m\frac{\rmd^2\bb_i\rmd^2\br_i}{\pi B_p}e^{-\frac{b_i^2}{B_p}}\delta^2(\br_i)\left<\prod_{i=1}^mD\left(\bb_i,\bb_i\right)\right>\nonumber
\\
&=&1,
\end{eqnarray}
where we used
\begin{eqnarray}
\left<\prod\limits_{i=1}^mD\left(\bb_i,\bb_i\right)\right>=\frac{1}{N_c^m}\left<\prod\limits_{i=1}^m\text{Tr}\left[U\left(\bb_i\right)U\left(\bb_i\right)^{\dagger}
\right]\right>=\frac{1}{N_c^m}\left<\prod\limits_{i=1}^m\text{Tr}\left[\mathbf{1}_{(N_c\times N_c)}\right]\right>=1.
\end{eqnarray}

Substituting the factors $L$'s and $F$'s in Eq. (\ref{M3}) into Eq. (\ref{DDDM}), one obtains the explicit expression for the 3-dipole correlator up to the $\frac{1}{N_c^4}$ order
\begin{eqnarray}
&&\left<D(\bx_1,\bx_2)D(\bx_3,\bx_4)D(\bx_5,\bx_6)\right>\Bigg|_{\text{to the}\frac{1}{N_c^4}\text{order}}\nonumber
\\
&=&T_{\text{3-dipole}}e^{2C_FL_{12,34,56}+\frac{F_{1234,1256,3456}}{N_c}}\Bigg\{\frac{1}{3}+\int\limits_0^1\rmd\xi_1\int\limits_0^{\xi_1}\rmd\xi_2 F_{1243}F_{1423}
e^{(\xi_1-\xi_2)N_cF_{1342}}+\int\limits_0^1\rmd\xi_1\int\limits_0^{\xi_1}\rmd\xi_2\int\limits_0^{\xi_2}\rmd\xi_3\int\limits_0^{\xi_3}\rmd\xi_4F_{1243}F_{1423}\nonumber
\\
&&\times e^{(\xi_1-\xi_2)N_cF_{1342}}\Big[F_{1243}F_{1423}e^{(\xi_3-\xi_4)N_cF_{1342}}+2F_{1265}F_{1625}e^{(\xi_3-\xi_4)N_cF_{1562}}\Big]
+2\int\limits_0^1\rmd\xi_1\int\limits_0^{\xi_1}\rmd\xi_2\int\limits_0^{\xi_2}\rmd\xi_3\int\limits_0^{\xi_3}\rmd\xi_4F_{1243}F_{1465}\nonumber
\\
&&\times e^{N_c(\xi_1F_{1342}+\xi_2F_{1564})}\Big[F_{1645}F_{1423}e^{N_c(\xi_3F_{1546}+\xi_4F_{1324})}+F_{3245}F_{1625}e^{N_c(\xi_3F_{3542}+\xi_4F_{1526})}+F_{1623}F_{3645}
e^{N_c(\xi_3F_{1326}+\xi_4F_{3546})}\Big]\nonumber
\\
&&+\int\limits_0^1\rmd\xi_1\frac{F_{1243}}{N_c}e^{\xi_1N_cF_{1342}}+\int\limits_0^1\rmd\xi_1\int\limits_0^{\xi_1}\rmd\xi_2\int\limits_0^{\xi_2}\rmd\xi_3\bigg[
\frac{F_{1243}F_{1423}}{N_c}e^{(\xi_1-\xi_2)N_cF_{1342}}\Big(F_{1243}e^{\xi_3N_cF_{1342}}+F_{1265}e^{\xi_3N_cF_{1562}}\Big)\nonumber
\\
&&+\frac{F_{1265}F_{1625}F_{1243}}{N_c}e^{(\xi_1-\xi_2)N_cF_{1562}+\xi_3N_cF_{1342}}\bigg]+2\int\limits_0^1\rmd\xi_1\int\limits_0^{\xi_1}\rmd\xi_2\int\limits_0^{\xi_2}
\rmd\xi_3\frac{F_{1243}F_{1465}}{N_c}e^{N_c(\xi_1F_{1342}+\xi_2F_{1564})}\Big[F_{1645}e^{\xi_3N_c F_{1546}}\nonumber
\\
&&+F_{3245}e^{\xi_3N_cF_{3542}}+F_{1623}e^{\xi_3N_cF_{1326}}\Big]+2\int\limits_0^1\rmd\xi_1\int\limits_0^{\xi_1}\rmd\xi_2\frac{F_{1243}F_{1465}}{N_c^2}
e^{N_c(\xi_1F_{1342}+\xi_2F_{1564})}\Bigg\}\nonumber
\\
&&+\text{The same expression with indices inside $L$ and $F$ facors replaced by: }\{12,34,56\}\rightarrow\{34,56,12\}\text{ and }\{56,12,34\},\nonumber
\\
\end{eqnarray}
where in order to make the expression shorter, each $L$ and $F$ factor absorbs a $\frac{\mu^2}{2}$, and the integral variables $\xi$ in Eq. (\ref{DDDM}) are replaced by
\begin{eqnarray}
\int\limits_0^1\rmd\xi_0\rmd\xi_1\rmd\xi_2\rmd\xi_3\delta(\xi_0+\xi_1+\xi_2+\xi_3-1)f(\xi_0,\xi_1,\xi_2,\xi_3)=\int\limits_0^1\rmd\xi_1\int\limits_0^{\xi_1}\rmd\xi_2
\int\limits_0^{\xi_2}\rmd\xi_3f(1-\xi_1,\xi_1-\xi_2,\xi_2-\xi_3,\xi_3).
\end{eqnarray}
Using the functional forms \cite{Dominguez:2008aa} of the color transition factors $F$'s and $L$'s as
\begin{eqnarray}
\mu^2(L_{ii}+L_{jj}-2L_{ij})=\frac{Q_s^2}{2C_F}(\bx_i-\bx_j)^2,\ \ \ \mu^2 F_{1234}=\frac{Q_s^2}{2C_F}(\bx_1-\bx_2)\cdot(\bx_3-\bx_4),\label{LF}
\end{eqnarray}
and replacing the coordinates by
\begin{eqnarray}
\bx_{1,2}=\bb_1\pm\frac{\br_1}{2},\ \ \ \bx_{3,4}=\bb_2\pm\frac{\br_2}{2},\ \ \ \bx_{5,6}=\bb_3\pm\frac{\br_3}{2},
\end{eqnarray}
one reaches the final expression for $\left<\prod\limits_{i=1}^3D\left(\bb_i+\frac{\br_i}{2},\bb_i-\frac{\br_i}{2}\right)\right>$ to the $\frac{1}{N_c^4}$ order. We substitute this correlator into the definition of $c_{12}$ and $c_{123}$ in Eq. (\ref{c3}), and integrate over the variables $\bp,\ \bb,\ \br$ by the following five steps:\\
(a) Integrate over $\phi$ (the azimuthal angular of $\bp$) by
\begin{eqnarray}
\int\limits_0^{2\pi}\rmd\phi e^{i(\bp\cdot\br+n\phi)}=2\pi i^nJ_n(pr)e^{in\theta};
\end{eqnarray}
\\
(b) Integrate the Gaussian type functions of $\bb$;\\
(c) We truncate the expansion series of $e^{\sharp Q_s^2\br_i\cdot\br_j}$ at order $N$ as $\sum\limits_{n=0}^N\frac{(\sharp Q_s^2\br_i\cdot\br_j)^n}{n!}$, and integrate over $r,\ p$ by \begin{eqnarray}
\int\limits_0^{\infty}r\rmd rp\rmd pe^{-ar^2}J_n(pr)r^m=\begin{cases}
\frac{\Gamma(\frac{m}{2})\frac{n}{2}}{a^{\frac{m}{2}}}, &n>0,
\\
\\
\frac{\Gamma(\frac{m}{2})\frac{n}{2}}{a^{\frac{m}{2}}(-1)^n}, &n<0;
\end{cases}
\end{eqnarray}
\\
(d)Integrate over $\theta_i$ (the azimuthal angular of $\br_i$) analytically, and integrate over $\xi$ numerically.\\

\begin{figure}[H]
\vskip0\linewidth
\centerline{
\includegraphics[width = .4\linewidth]{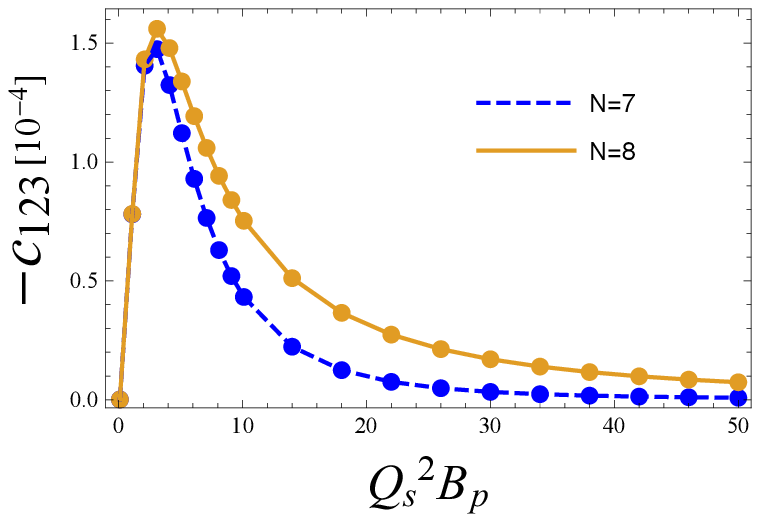}
\includegraphics[width = .42\linewidth]{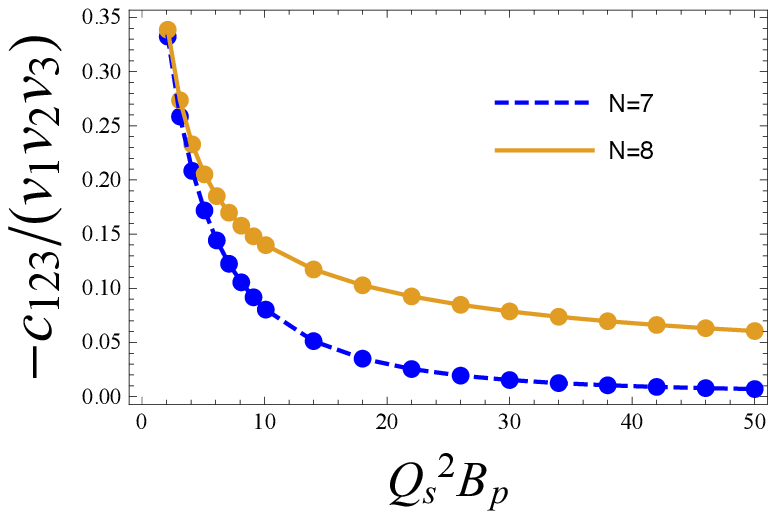}
}
\caption{Left: $-c_{123}$ depending on $Q_s^2B_p$, where we truncate the expansion series at orders seven and eight. We see in the Taylor expansion approach, $-c_{123}$ converges quickly in small $Q_s^2B_p$ region, and gradually converges in large $Q_s^2B_p$ region. Right: The scaled cumulant $-c_{123}$ as a function of $Q_s^2B_p$, where the factors in the denominator are written by $v_n\equiv v_n\{2\}$ for short.}
\label{c123}
\end{figure}

Then we reach the values of the 3-particle cumulant $c_{123}$ (with negative values) up to the $\frac{1}{N_c^4}$ order in the MV model, which is plotted as a function of $Q_s^2B_p$ in Fig. \ref{c123}. Some 3-particle charge-dependent azimuthal correlations have been measured at RHIC and the LHC \cite{Abelev:2009ac,Khachatryan:2016got} to search for the chiral magnetic effect. We see that our $-c_{123}$ (with a falloff at large $Q_s^2$) has a similar magnitude and trend to the 3-particle correlations $\left<\text{cos}(\phi_\alpha+\phi_\beta-2\phi_c)\right>$ (whose magnitude gradually decreases as the event mutiplicity increases) observed in pPb collisions at the CMS \cite{Khachatryan:2016got}, where $Q_s$ is related to the multiplicity (centrality) by growing (falling) with the multiplicity (centrality). In order to obtain a number of order unity, we scale the mixed cumulant $c_{123}$ by the corresponding 2-particle anisotropic flows $v_n\{2\}\equiv\sqrt{c_n\{2\}}$ \cite{Davy:2018hsl}, and plot the ratios as in Fig. \ref{c123}, which also shows a similar magnitude and trend to the result in Ref. \cite{Bhalerao:2011yg}.

\section{The $N_c$ scaling of multiparticle correlations}\label{scaling}
For the $2m$-particle harmonic, one usually defines
\begin{eqnarray}
\kappa_n\{2m\}\equiv\int\prod_{i=1}^m\rmd^2\bp_ie^{in(\phi_1+...+\phi_m-\phi_{m+1}-...-\phi_{2m})}\frac{\rmd^mN}{\prod_{i=1}^m\rmd^2\bp_i},\ \ \ c_n\{2m\}\equiv\frac{\kappa_n\{2m\}}{\kappa_0\{2m\}},
\end{eqnarray}
for convenience.

\begin{figure}[H]
\vskip0.0\linewidth
\centerline{
\includegraphics[width = .6\linewidth]{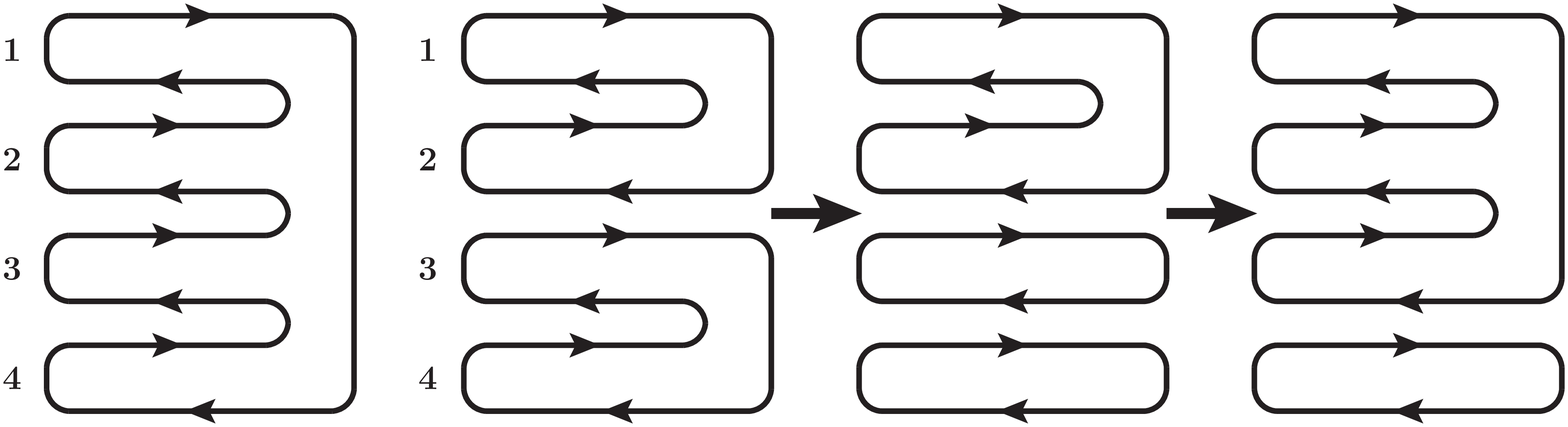}
}
\caption{The left topology represents that all the four dipoles connect together. The right topology represents that dipoles $3,4$ change from connected to disconnected, and finally dipoles $1,2,3$ connect together. Both of them are of $\frac{1}{N_c^6}$ order.}
\label{4connect}
\end{figure}

Since the $m$-particle cumulant $c_n\{m\}$ demonstrates the correlations of these $m$ particles, as a prior guest,  its $N_c$ scaling should depend on the configurations where all dipoles (particles) connect or ever connected together as in Fig. \ref{4connect}.

As each gluon exchanging between two unconnected dipoles brings a $\frac{1}{N_c^2}$, the cumulants should scale as \cite{Fukushima:2017mko}
\begin{eqnarray}
c_n\{m\}\sim\frac{1}{N_c^{2(m-1)}},\label{Ncscale}
\end{eqnarray}
which is well understood graphicly, and this can also be proved by the matrix method. We consider the $\frac{1}{N_c^4}$ order of $c_n\{4\}\equiv\frac{\kappa_n\{4\}}{\kappa_0\{4\}}-2\left(\frac{\kappa_n\{2\}}{\kappa_0\{2\}}\right)^2=\left<e^{in(\phi_1-\phi_2+\phi_3-\phi_4)}\right>-2\left<e^{in(\phi_1-\phi_2)}\right>^2$, which means only two gluons exchanging between dipoles $(1,2)$, $(3,4)$ or $(1,4)$, $(3,2)$ contribute, and thus we only use these two 2-gluon exchanging matrices instead of the full 4-dipole color transition matrix as
\begin{eqnarray}
M_{(1,2),(3,4)}&=&
\mu^2\left(\begin{matrix}LF_{12,34;56,78}&\frac{1}{2}F_{5867}&\frac{1}{2}F_{1423}&0
\\\frac{1}{2}F_{5687}&LF_{12,34;58,76}&0&\frac{1}{2}F_{1423}
\\\frac{1}{2}F_{1243}&0&LF_{14,32;56,78}&\frac{1}{2}F_{5867}
\\0&\frac{1}{2}F_{1243}&\frac{1}{2}F_{5687}&LF_{14,32;58,76}\end{matrix}\right)\nonumber
\\
&=&\mu^2\left(\begin{matrix}LF_{12,34}&\frac{1}{2}F_{1423}\\\frac{1}{2}F_{1243}&LF_{14,32}\end{matrix}\right)\otimes\mathbf{1}
+\mathbf{1}\otimes\mu^2\left(\begin{matrix}LF_{56,78}&\frac{1}{2}F_{5687}\\\frac{1}{2}F_{5867}&LF_{58,76}\end{matrix}\right)\nonumber
\\
&=&M_{(1,2)}\otimes\mathbf{1}+\mathbf{1}\otimes M_{(3,4)}\nonumber
\\
&\equiv&M_{(1,2)}\oplus M_{(3,4)},\label{decompose}
\end{eqnarray}
and $M_{(1,4),(3,2)}=M_{(1,4)}\oplus M_{(3,2)}$ in the same way, where $LF_{ab,cd;ef,gh}\equiv LF_{ab,cd}+LF_{ef,gh}$. Thus we see that $M_{(1,2),(3,4)}$  can be decomposed into the Kronecker sum of two individual 2-dipole color transition matrices $M_{(1,2)}$ and $M_{(3,4)}$, and so can $M_{(1,4),(3,2)}$. Then the first term of $c_n\{4\}$ gives
\begin{eqnarray}
\left<e^{in(\phi_1-\phi_2+\phi_3-\phi_4)}\right>\Big|_{\text{to the}\frac{1}{N_c^4}\text{order}}
&=&\int{\prod_{i=1}^4}\frac{\rmd^2\bb_i\rmd^2\br_i\rmd^2\bp_i}{4\pi^3B_p}e^{-\frac{b_i^2}{B_p}-\frac{r_i^2}{4B_p}+i\bp_i\cdot \br_i}
e^{in(\phi_1-\phi_2+\phi_3-\phi_4)}\nonumber
\\
&&\times\left(\begin{matrix}1&\frac{1}{N_c}&\frac{1}{N_c}&\frac{1}{N_c^2}\end{matrix}\right)
\left[e^{M_{(1,2),(3,4)}}+e^{M_{(1,4),(3,2)}}\right]\left(\begin{matrix}1\\0\\0\\0\end{matrix}\right)\Bigg|_{\text{to the}\frac{1}{N_c^4}\text{order}},\label{<1234>}
\end{eqnarray}
where
\begin{eqnarray}
&&\left(\begin{matrix}1&\frac{1}{N_c}&\frac{1}{N_c}&\frac{1}{N_c^2}\end{matrix}\right)=\left(\begin{matrix}1&\frac{1}{N_c}\end{matrix}\right)
\otimes\left(\begin{matrix}1&\frac{1}{N_c}\end{matrix}\right),
\\
&&e^{M_{(1,2),(3,4)}}=e^{M_{(1,2)}\oplus M_{(3,4)}}=e^{M_{(1,2)}}\otimes e^{M_{(3,4)}},
\\
&&\left(\begin{matrix}1\\0\\0\\0\end{matrix}\right)=\left(\begin{matrix}1\\0\end{matrix}\right)\otimes\left(\begin{matrix}1\\0\end{matrix}\right),
\end{eqnarray}
which gives
\begin{eqnarray}
\left(\begin{matrix}1&\frac{1}{N_c}&\frac{1}{N_c}&\frac{1}{N_c^2}\end{matrix}\right)e^{M_{(1,2),(3,4)}}\left(\begin{matrix}1\\0\\0\\0\end{matrix}\right)
&=&\left(\begin{matrix}1&\frac{1}{N_c}\end{matrix}\right)e^{M_{(1,2)}}\left(\begin{matrix}1\\0\end{matrix}\right)
\otimes\left(\begin{matrix}1&\frac{1}{N_c}\end{matrix}\right)e^{M_{(3,4)}}\left(\begin{matrix}1\\0\end{matrix}\right)\nonumber
\\
&=&\left<D(\bb_1+\frac{\br_1}{2})D(\bb_2-\frac{\br_2}{2})\right>\left<D(\bb_3+\frac{\br_3}{2})D(\bb_4-\frac{\br_4}{2})\right>.\label{<><>}
\end{eqnarray}

Then substituting Eq. (\ref{<><>}) into Eq. (\ref{<1234>}) gives
\begin{eqnarray}
\left<e^{in(\phi_1-\phi_2+\phi_3-\phi_4)}\right>\Big|_{\text{to the}\frac{1}{N_c^4}\text{order}}
&=&\left<e^{in(\phi_1-\phi_2)}\right>\left<e^{in(\phi_3-\phi_4)}\right>+\left<e^{in(\phi_1-\phi_4)}\right>\left<e^{in(\phi_3-\phi_2)}\right>\Big|_{\text{to the}\frac{1}{N_c^4}\text{order}}\nonumber
\\
&=&2\left<e^{in(\phi_1-\phi_2)}\right>^2\Big|_{\text{to the}\frac{1}{N_c^4}\text{order}},
\end{eqnarray}
which proves that $c_n\{4\}$ up to the $\frac{1}{N_c^4}$ order is 0. Nevertheless, the $\frac{1}{N_c^6}$ order involves new configurations that all the four dipoles connect with each other, and they cannot be subtracted by term $2\left<e^{in(\phi_1-\phi_2)}\right>^2$ in the definition of $c_n\{4\}$.

By the same way, for the $c_n\{m\}$ case, the configurations that can be divided into unconnected parts (and the corresponding color transition matrices can be decomposed into the Kronecker sum of individual matrices as Eq. \ref{decompose}) in $\frac{\kappa_n\{m\}}{\kappa_0\{m\}}$, is totally subtracted by the minus terms in $c_n\{m\}$. Therefore the conclusion of the $N_c$ scaling in Eq. (\ref{Ncscale}) is proved.

Finally, we can calculate the small $Qs$ expansions of $c_2\{2\}$ and $c_2\{4\}$ with the color transition matrices $M_{2-\text{dipole}}$ and $M_{4-\text{dipole}}$. With a conjectured $2m$-particle cumulant formula
\begin{eqnarray}
c_n\{2m\}\equiv\frac{\kappa_n\{2m\}}{\kappa_0\{2m\}}-\sum_{j=2}^m\left\{\left[\prod_{i=1}^{j-1}\sum_{a_i=1}^{a_{i+1}-1}\right]\left[\frac{1}{j!}\prod_{i=1}^jc_n\{2(a_i-a_{i-1})\}\binom{a_i- a_{i-1}}{a_i}^2\right]\Bigg|_{a_0=0,a_j=m}\right\},
\end{eqnarray}
one obtains the definitions \cite{Borghini:2001vi}
\begin{eqnarray}
c_n\{2\}\equiv\frac{\kappa_n\{2\}}{\kappa_0\{2\}},\ \ c_n\{4\}\equiv\frac{\kappa_n\{4\}}{\kappa_0\{4\}}-2\left(\frac{\kappa_n\{2\}}{\kappa_0\{2\}}\right)^2,\ \ c_n\{6\}\equiv\frac{\kappa_n\{6\}}{\kappa_0\{6\}}-9\frac{\kappa_n\{2\}}{\kappa_0\{2\}}\frac{\kappa_n\{4\}}{\kappa_0\{4\}}+12\left(\frac{\kappa_n\{2\}}{\kappa_0\{2\}}\right)^3,....
\end{eqnarray}

Substituting $M_{2-\text{dipole}}$ and $M_{4-\text{dipole}}$ into the above definitions gives the small $Qs$ expansions in the leading $N_c$ order as
\begin{eqnarray}
&&c_2\{2\}=\frac{1}{N_c^2}\left[\frac{1}{2}Q_s^4B_p^2-Q_s^6B_p^3+\frac{23}{16}Q_s^8B_p^4-\frac{437}{240}Q_s^{10}B_p^5+\mathcal{O}\left(Q_s^{12}B_p^6\right)\right]+\mathcal{O}\left(\frac{1}{N_c^4}\right),\nonumber
\\
&&c_2\{4\}=\frac{1}{N_c^6}\left[Q_s^8B_p^4-\frac{20}{3}Q_s^{10}B_p^5+\frac{1781}{72}Q_s^{12}B_p^6-\frac{686351}{10080}Q_s^{14}B_p^7+\mathcal{O}\left(Q_s^{16}B_p^8\right)\right]+\mathcal{O}\left(\frac{1}{N_c^8}\right),
\end{eqnarray}
as well as the 3-particle cumulants:
\begin{eqnarray}
c_{123}=\frac{-\pi}{N_c^4}\left[\frac{3}{64}Q_s^8B_p^4-\frac{109}{512}Q_s^{10}B_p^5+\frac{7091}{12288}Q_s^{12}B_p^6+\mathcal{O}\left(Q_s^{14}B_p^7\right)\right]+\mathcal{O}\left(\frac{1}{N_c^6}\right),
\end{eqnarray}
which finally helps to crosscheck the $N_c$ scaling of $c_n\{m\}$ in Eq. (\ref{Ncscale}).

\section{Conclusion}\label{Conclusion}
In this paper, we evaluated the 3-dipole correlator up to the $\frac{1}{N_c^4}$ order in the CGC formalism, and conjectured an $n$-dipole formula as a $\frac{1}{N_c}$ power expansion. A clear pattern was observed that each term of the $n$-dipole correlator corresponds to individual color transitions between configurations. We calculated the azimuthal cumulant $c_{123}$ by using the 3-dipole correlator, and proved the $N_c$ scaling of $c_n\{m\}$, which was cross-checked by our calculation of $c_2\{2\}$, $c_2\{4\}$ and $c_{123}$ in the dilute limit.

\begin{acknowledgments}
We thank Dr. Bo-Wen Xiao for useful discussions and comments. This material is based on the work supported by the Natural Science Foundation of China (NSFC) under Grants No. 11575070, No. 11435004, and No. 11805167.
\end{acknowledgments}

\end{document}